\documentclass[a4paper,fleqn]{cas-sc}

\usepackage[numbers]{natbib}
\usepackage{algorithm}
\usepackage{algpseudocode}
\usepackage{xcolor}
\usepackage{soul}
\usepackage{graphicx}
\usepackage{subcaption}
\usepackage{graphicx}
\newtheorem{claims}{Claim}

\def\tsc#1{\csdef{#1}{\textsc{\lowercase{#1}}\xspace}}
\tsc{WGM}
\tsc{QE}
\tsc{EP}
\tsc{PMS}
\tsc{BEC}
\tsc{DE}

\begin{document}
\let\WriteBookmarks\relax
\def\floatpagepagefraction{1}
\def\textpagefraction{.001}
\shorttitle{Fixed Order Scheduling with Deadlines}
\shortauthors{A. Berger, A. Rouhani, and M. Schr{\"o}der}

\title [mode = title]{Fixed Order Scheduling with Deadlines}                      



\author[1]{Andr{\'e} Berger}[
                        auid=000,bioid=1,
                        orcid=0000-0002-6409-1963]
\ead{a.berger@maastricjhtuniversity.nl}

\author[1]{Arman Rouhani}[type=editor,
                        auid=000,bioid=1,
                        orcid=0000-0003-4822-484X]
\cormark[1]
\ead{a.rouhani@maastricjhtuniversity.nl}
\author[1]{Marc Schr{\"o}der}[type=editor,
                        auid=000,bioid=1,
                        orcid=0000-0002-0048-2826]
\ead{m.schroder@maastricjhtuniversity.nl}

\affiliation[1]{organization={Department of Quantitative Economics, Maastricht University},
                addressline={Tongersestraat 53}, 
                postcode={6211 LM}, 
                postcodesep={}, 
                city={Maastricht},
                country={The Netherlands}}






 \cortext[cor1]{Corresponding author}


\begin{abstract}
This paper studies a scheduling problem where machines must follow a predetermined fixed order for processing jobs. Given $n$ jobs, each with processing times and deadlines, we aim to minimize the number of machines used while meeting deadlines and maintaining the order. We show that the first-fit algorithm is optimal for unit processing times and is a 2-approximation when the order aligns with non-increasing slacks, non-decreasing slacks, or non-increasing deadlines. For the general order, we provide an $O(\log n)$-approximation.
\end{abstract}




\maketitle

\section{Introduction}
Scheduling problems are a well-studied field in the domain of operations research. Given a set of jobs~$J$, where each job \( j \in J \) has a release date \( r_j \), a processing time \( p_j \), and a deadline \( d_j \), along with a set of machines, the goal is to determine a feasible assignment of jobs to machines that optimizes the objective of the problem while respecting the constraints on the job details and the machine environment. In the literature, various objectives such as makespan, completion time, number of late jobs are examined. Among these objectives, \textit{minimizing the number of machines} is very important in practice. 

Consider a practical scenario where multiple users utilize a set of parallel processors with shared memory. By scheduling jobs on a certain subset of processors, processors can be put to sleep, thereby reducing energy consumption. The objective is to minimize the cost, specifically the number of processors in use. In numerous practical applications such as airport gate management \cite{kroon1997exact}, train scheduling \cite{eidenbenz2003flexible}, the transportation of navy fuel oil tankers \cite{dantzig2003minimizing}, developing distributed memory architectures \cite{moukrim1999minimum}, runway scheduling, and scheduling of maintenance work for trains in a service station, the processing of a set of jobs on a minimum number of machines is desired while the schedule respects the time intervals in which jobs have to be scheduled. These problems are referred to as \textit{SRDM} (scheduling with release times and deadlines on a minimum number of machines \cite{cieliebak2004scheduling}), and in the scheduling notation $\alpha|\beta|\gamma$ introduced by Graham
et al. \cite{graham1979optimization}, it is represented as $P|r_j,p_j, d_j|m$. For the special case of the problem where the jobs have a common release time and equal deadlines, the problem is reduced to the well-known bin packing problem (BPP), which is known to be NP-hard \cite{coffman1997approximation}. The general problem can be considered a BPP with additional constraints. Specifically, when $d_j = r_j + p_j$ for each job $j \in J$, the problem becomes a BPP with conflicts \cite{delorme2016bin,pereira2016procedures,sadykov2013bin}. Cieliebak et al. \cite{cieliebak2004scheduling} showed that when $d_j \leq r_j + p_j + 1$, the SRDM problem can be solved in polynomial time; otherwise, it remains NP-complete. For the latter case, they also provide approximation algorithms. Yu and Zhang \cite{yu2009scheduling} provided a 2-approximation algorithm for the common release time variant of the problem denoted as $P|r_j = r,p_j, d_j|m$. Moreover, they proved that the problem of $P|r_j,p_j = p, d_j|m$ can be 6-approximated. This result was later improved by Kravchenko and Werner \cite{kravchenko2009minimizing}, who proved that the equal processing times variant of the problem is polynomially solvable.

In various practical applications such as CPU task management, perishable goods handling, airline boarding, and healthcare systems, there is often a requirement to follow an \textit{order} on each machine when scheduling tasks. This fixed order is distinct from precedence constraints, as it requires tasks to be processed in a specific sequence on each processor, without dependencies between the tasks themselves. For example, in CPU scheduling with static priorities, tasks must be executed in a pre-established sequence across processors, following global priority rules \cite{buttazzo1997hard}. Similarly, in industries dealing with perishable goods, warehouses or delivery systems must follow a predefined order to ensure items are processed and delivered before their expiration \cite{nahmias1982perishable}. In airline operations, passengers are boarded in a fixed sequence across gates, ensuring smooth and efficient boarding procedures \cite{bazargan2007linear}. In healthcare systems, patient appointments are frequently managed in a fixed sequence, with doctors treating patients in the order determined by their schedule, independent of other appointments \cite{gupta2008appointment}. These examples demonstrate the importance of efficient scheduling under a fixed global order. This variant of the problem recently gained attention for different objectives \cite{bosman2019fixed,vijayalakshmi2024minimizing}.

In this paper, we address the SRDM problem with common release times and a fixed processing order, denoted as $P|r_j=r, p_j, d_j, \pi^{*}|m$. We investigate the impact of the imposed processing order on job scheduling and study the approximation factor of some of the simplest greedy algorithms for the problem: first-fit and next-fit.

Section 2 starts by formally defining the problem and introducing the notations used throughout the paper. In Section 3, we show that first-fit is an optimal algorithm when the processing times are equal and then show it is a 2-approximation for several variants of the problem. Lastly, we give a $O(\log n)$ approximation problem for the general problem. Finally, in Section 4, we summarize our findings, emphasize our contributions, and suggest directions for future research.

\section{Preliminaries}
In this study, our focus lies on determining the minimum number of machines required to schedule a given set of jobs. The input to the problem consists of a set of $n$ jobs denoted as $J = \{1,\ldots,n \}$. Additionally, we have a sufficiently large supply of identical machines $\{1,\ldots,n\}$ at our disposal. Each job $j \in J$ is characterized by a pair $(p_j,d_j)\in  \mathbb{N}^2$ with processing time $p_j > 0$, reflecting the time required for completion, and deadline $d_j > 0$, where $d_j \geq p_j$. For each job $j \in J$, we define $\lambda_j = d_j - p_j$ as the $\textit{slack}$ representing the difference between the job's deadline and its processing time. Let $T = \{(p_j, d_j)\mid j\in J\}$ be the set of tuples representing the processing time and deadline for each job $j \in J$. The jobs follow a predetermined fixed order for processing, and each machine schedules its assigned jobs according to this order. W.l.o.g., we assume that job 1 has the highest priority, job 2 has the second-highest priority, and job $n$ has the lowest priority, i.e., $j<k$ means that $j$ has a higher priority than $k$. The objective of the problem is to minimize the number of machines necessary for scheduling the jobs while ensuring that the prescribed order is maintained on each machine, and every job is completed within its stipulated deadline. Given that all jobs have the same release time of 0, we assume that there are no idle times between jobs and moreover, preemption of processing is not allowed, i.e. the processing of any job $j$ started at time $t$ on one of the machines will be completed at time $p_j + t$ on the same machine and each machine can only perform one job at a time. 

An optimal solution is a schedule $\tau^*:J\rightarrow \{1,\ldots,n\}$ that is feasible, i.e., $\sum_{k\leq j: \tau^*(k)=\tau^*(j)}p_k\leq d_j$ for all $j\in J$, and minimizes the number of open machines, i.e., a machine is \textit{open} if it has been assigned at least one job. Let $\mathcal{M^*}$ denote the set of open machines. For an optimal schedule (\textit{OPT}), define $J^*_i=\{j\in J\mid \tau^*(j)=i\}$ as the set of jobs assigned to machine $i\in \mathcal{M^*}$, and $L^*_i=\sum_{j\in J^*_i}p_j$ as the load of machine $i\in\mathcal{M^*}$. Moreover, let $l^*_i$ denote the last job scheduled on machine $i \in \mathcal{M^*}$. We will study two simple greedy algorithms. The \textit{first-fit (FF)} algorithm schedules jobs based on the fixed order and assigns each job to the first feasible machine. An even simpler greedy algorithm \textit{next-fit} (NF) operates similarly with the exception that NF only considers the last open machine (while FF considers all open machines) for scheduling a job and, if that job fails to fit in the last machine NF uses a new empty machine for scheduling. Given some approximation algorithm with schedule $\tau:J\rightarrow \{1,\ldots,n\}$, we define $J_i=\{j\in J\mid \tau(j)=i\}$ as the set of jobs assigned to machine $i$. Furthermore, let $L_i=\sum_{j\in J_i}p_j$ represent the load of machine $i$. 
Let $OPT(I)$ denote the number of open machines of an optimal solution and $ALG(I)$ denote the number of open machines of the algorithm for a given instance $I$. The approximation ratio of an algorithm is defined as $\sup_I\frac{ALG(I)}{OPT(I)}$. In this paper, we will study the approximation ratio of first-fit and next-fit for the fixed order scheduling problem.

\section{Analysis of Approximation Factors}

In this section, we provide a comprehensive analysis of our problem. Initially, we show that there does not exist a constant approximation factor for next-fit. Next, we establish the optimality of first-fit in the case of unit processing times. Then, we prove that first-fit has an approximation factor of 2 in four different classes of instances: (1) if the fixed order aligns with slacks that are in a non-increasing order, (2) if the fixed order aligns with slacks that are in a non-decreasing order, (3) if the fixed order aligns with deadlines that are in a non-increasing order, and (4) if the optimal solution uses at most 3 machines. For the case of non-decreasing deadlines, an approximation factor dependent on the deadline and processing time of the last job is provided.

\textbf{Instance 1.} Let $\mathcal{I}_n$ be a parametric instance with $n > 2$ jobs. Assume that $(p_1,d_1)=(1,1)$ and $(p_2,d_2)=(2,2)$. For $j>2$, the pair $(p_j, d_j)$ is obtained by $(p_{j-1} + p_{j-2}, p_j + p_{j-1} - 1)$. 
\begin{figure}[t!]
\centering
\includegraphics[scale=1]{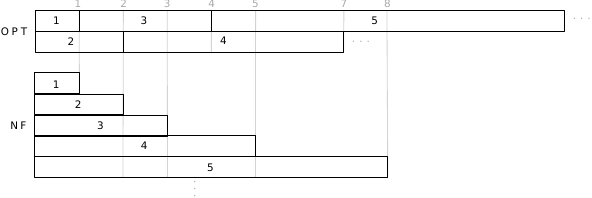}
\caption{The optimal schedule and the schedule by NF for $\mathcal{I}_n$ where $n\geq 5$.} \label{NFunbounded}
\end{figure}
\begin{lemma}
    Next-fit has an unbounded approximation ratio for the instance $\mathcal{I}_n$.
\end{lemma}
\begin{pf}
The NF algorithm consistently selects the most recently opened machine to place a new job $j$. When applied to $\mathcal{I}_n$, the NF algorithm initiates the opening of a new machine for each job $j \in J$ since the deadline of job $j$ is equal to $p_j + p_{j-1} - 1$. However, the optimal schedule requires only two machines to schedule all the jobs. Jobs with odd indices are scheduled on one machine, and jobs with even indices are scheduled on the other machine (see Figure \ref{NFunbounded} for an illustration). Hence, the number of machines in OPT is always $2$ and the number of machines in the NF algorithm is equal to $n$. Thus, the lemma holds.\qed
\end{pf}

\subsection{Unit Processing Time}
In this subsection, we consider the first-fit algorithm for the case of unit processing times. In scheduling notation, this problem is represented as $P|r_j = r,p_j = 1, d_j, \pi^{*}|m$.

\begin{theorem}
    The FF algorithm solves the problem $P|r_j = r,p_j = 1, d_j, \pi^{*}|m$ optimally in time $O(n^2)$.
\end{theorem}
\begin{pf}
    It is easy to verify that the worst-case running time of the FF algorithm is $O(n^2)$. Therefore, it only remains to prove that the FF algorithm provides an optimal schedule for $P|r_j = r,p_j = 1, d_j, \pi^{*}|m$.

    Denote by $\tau$ the schedule produced by the FF algorithm. Define $J^{\pi^{*}}_k=\{1,\ldots,k\}$ and let $\tau_{k}$ be the schedule produced by the FF algorithm for the jobs $J^{\pi^{*}}_k.$ Furthermore, we call the position of a job \(j\) on a machine its \textit{spot} $s$, i.e., the spot of job $j\in J$ is $s(j)=|\{k\in J\mid k\leq j\text{ and }\tau(k)=\tau(j)\}|$.

    \begin{claims}\label{lm:sorted-loads}
    If the jobs have unit processing times, then applying the FF algorithm yields
    \begin{equation}
        L_1 \geq L_2 \geq \ldots \geq L_n.
    \end{equation}
    \end{claims}
    \begin{pf}
    Assume by contradiction that the load of a machine $i$ is less than the load of a machine $i^{'}$, where $i<i^{'}$. Now, consider the iteration of the FF algorithm when $L_{i^{'}}$ becomes larger than $L_i$ which means $L_{i^{'}} = L_i + 1$. Observe that since the processing times are unit, before the iteration both of the machines have equal loads. This means that since the job fits in $i^{'}$ it could have fit in machine $i$. This is a contradiction to the procedure of the FF algorithm since it fits a job in a machine with the lowest possible index.
    \qed
    \end{pf}
    \begin{claims}
        There exists an optimal solution for the problem $P|r_j = r,p_j = 1, d_j, \pi^{*}|m$ where the assignment of jobs of $J^{\pi^{*}}_k$ are the same as their assignment in $\tau_{k}$, for all $1\leq k \leq n$. 
    \end{claims}
    \begin{pf}
        The proof is done by induction on $k$.
        
        \textbf{Base Case:  }By the assignment of the first job, the load of one machine becomes one and the rest have load zero which is not dependent on the assignment. Therefore, it is easy to verify that the base case is valid.

        \textbf{Induction Hypothesis:  }We assume that there exists an optimal schedule $\tau^{*}$ that the assignment of each $j \in J^{\pi^{*}}_{k-1}$ is the same as in $\tau$.

        \textbf{Induction Step:  } We prove that there exist an optimal schedule $\tau^{**}$ such that the assignment of job $k$ is the same as in $\tau$. Therefore, we modify the schedule of $\tau^{*}$ that agrees with the schedule of $\tau$ for every $j \in J^{\pi^{*}}_{k-1}$ in a way that both $\tau$ and $\tau^{**}$ have the same placement for job $k$. Let $\tau(k)$ and $\tau^{*}(k)$ be the machines that job $k$ is scheduled in the FF algorithm and the optimal schedule, respectively. If $\tau(k)=\tau^{*}(k)$, then we are done because we can define $\tau^{**}=\tau^{*}$. So, we can assume that $\tau(k) \neq \tau^{*}(k)$. Because the FF algorithm assigns the jobs to the first open machine, we have $\tau(k) < \tau^{*}(k)$.
        
        Assume that job $k$ is assigned to spot $s$ on machine $\tau(k)$, and to spot $s^*$ on machine $\tau^{*}(k)$. Since the assignment of each $j \in J^{\pi^{*}}_{k-1}$ is the same as in $\tau$ and $\tau(k) < \tau^{*}(k)$, we have by Claim \ref{lm:sorted-loads} that $s\geq s^*$. Moreover, let $U_{\tau(k)}$ be the set of jobs $j \in J$ where $\tau^{*} (j) = \tau(k)$ and $s(j)>s$ (note that these jobs have lower priority than job $k$), and denote by $U_{\tau^{*}(k)}$ as the set of jobs $j \in J$ with $\tau^{*} (j) = \tau^{*} (k)$ and $s(j)>s^*$. Let also $k^{\prime}$ be the job that is assigned to the spot $s$ of the machine $\tau(k)$ in $\tau^{*}$ (note that this job need not exist). 
        
        First, assume that $s^* = s$. In this case, define $\tau^{**}$ as $\tau^*$ subject to the following change: process jobs in $U_{\tau(k)}$ on $\tau^{*}(k)$ and jobs in $U_{\tau^{*}(k)}$ on $\tau(k)$. By doing so, each job keeps the previous spot in the new machine and the order is not violated. Then, swap jobs $k$ and $k^{\prime}$ so that job $k$ is assigned to the machine $\tau(k)$ in $\tau^{**}$ (see Figure \ref{fig:s=s} for an illustration). This means there exists an optimal schedule $\tau^{**}$ that has the same assignment for job $k$ as in $\tau$.
         \begin{figure}[t!]
          \centering
          \begin{subfigure}{0.40\textwidth}
            \centering
            \includegraphics[width=\textwidth]{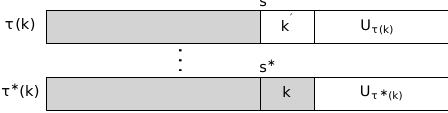}
            \caption{Assignment of jobs before modification.}
            \label{fig:sub1s=s}
          \end{subfigure}
          \hspace{0.05cm}
          \begin{subfigure}{0.40\textwidth}
            \centering
            \includegraphics[width=\textwidth]{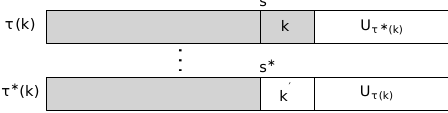}
            \caption{Assignment of jobs after modification.}
            \label{fig:sub2s=s}
          \end{subfigure}
          \caption{Exchanging the jobs between two machines of $\tau^*$ when $s = s^*$.}
          \label{fig:s=s}
        \end{figure}  
        
        Next, assume that $s^* < s$. We define $\tau^{**}$ by modifying $\tau^{*}$ in a way that it has the same placement for job $k$ as in $\tau$. If no job is scheduled in spot $s$ of $\tau(k)$, then job $k$ can be relocated to this machine in spot $s$, and we are done. Denote by $U^p_{\tau^{*}(k)}$ and $U^s_{\tau^{*}(k)}$ the set of jobs in $U_{\tau^{*}(k)}$ that precede and succeed job $k^{\prime}$ in $\pi^{*}$. Since job $k$ can be assigned to machine $\tau(k)$ in spot $s$ in $\tau$ and based on the induction hypothesis, changing the machine of job $k$ in $\tau^{*}$ does not violate its deadline. Moreover, the completion times of jobs in $U_{\tau^{*}(k)}$ are reduced. Based on the difference of the $s$ and $s^*$, we analyze two different cases
        
        \textbf{\textit{Case (I):} $|U^p_{\tau^{*}(k)}| \leq s - s^*$ -} In this case, define $\tau^{**}$ so that the assignments of jobs $k$ and $k^{\prime}$ are switched in a way that job $k$ is scheduled in spot $s$ of $\tau(k)$ and job $k^{\prime}$ is scheduled after $U^p_{\tau^{*}(k)}$ in the machine  $\tau^{*}(k)$ (see Figure \ref{fig:sub1s<s} for an illustration). It is easy to confirm that this modification violates no deadlines and preserves the order. Only the completion time of job $k$ increases and since it is assigned to spot $s$ of machine $\tau(k)$ in $\tau$, this swap does not violate its deadline. It is also worth mentioning that since $|U^p_{\tau^{*}(k)}| \leq s - s^*$, the deadline of job $k^{\prime}$ is not violated either. Thus, the proof holds.
        
        \textbf{\textit{Case (II):} $|U^p_{\tau^{*}(k)}| > s - s^*$-} In the previous case, since $|U^p_{\tau^{*}(k)}| \leq s - s^*$, switching the machine of $k^{\prime}$ to $\tau^{*}(k)$ does not violate the deadline of $k^{\prime}$. However, this becomes an issue in the current case. To solve this issue, we divide the jobs in $U^p_{\tau^{*}(k)}$ into two sets $U^{p1}_{\tau^{*}(k)}$ and $U^{p2}_{\tau^{*}(k)}$ where the former set consists of the first $s - s^*$ jobs of $U^p_{\tau^{*}(k)}$ and the latter set contains the rest. In the modified schedule $\tau^{**}$ (refer to Figure \ref{fig:sub2s<s} for an illustration) job $k$ is assigned to spot $s$ of $\tau(k)$ and job $k^{\prime}$ is assigned to spot $s$ of $\tau^{*}(k)$ while their orders and completion times are preserved. Jobs of $U^{p1}_{\tau^{*}(k)}$ are placed in the spots $s^*$ to $s - 1$ of machine $\tau^{*}(k)$ where their completion times are reduced and their orders are preserved. Jobs belonging to $U^{p2}_{\tau^{*}(k)}$ and $U_{\tau(k)}$ are assigned after spot $s$ in $\tau(k)$, as in the original configuration (before modification). Also, it is easy to verify that jobs $U^s_{\tau^{*}(k)}$ preserve their order after modification and their deadlines are not violated.
            \begin{figure}[t!]
              \centering
              \begin{subfigure}{0.42\textwidth}
                \centering
                \includegraphics[width=\textwidth]{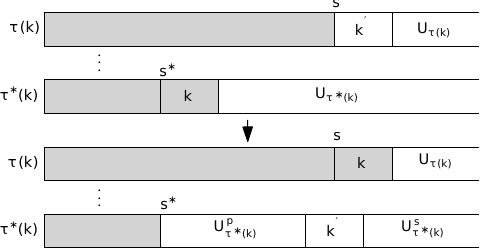}
                \caption{Assignment of jobs for $|U^p_{\tau^{*}(k)}| \leq s - s^*$.}
                \label{fig:sub1s<s}
              \end{subfigure}
              \hspace{0cm}
              \begin{subfigure}{0.42\textwidth}
                \centering
                \includegraphics[width=\textwidth]{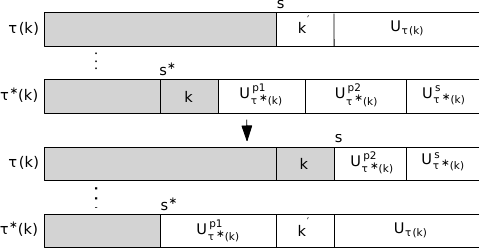}
                \caption{Assignment of jobs for $|U^p_{\tau^{*}(k)}| > s - s^*$.}
                \label{fig:sub2s<s}
              \end{subfigure}
              \caption{Exchanging the jobs between two machines of $\tau^*$ when $s > s^*$.}
              \label{fig:s<s}
            \end{figure}
            \qed
    \end{pf}
    Hence, FF provides an optimal schedule for $P|r_j = r,p_j = 1, d_j,  \pi^{*}|m$.
\qed
\end{pf}

\subsection{Non-Increasing Slacks}

In this section, we first prove that when the fixed order satisfies the non-increasing slacks property, the output of the next-fit algorithm is identical to the first-fit algorithm. In this case, the approximation ratio of both algorithms is equal to 2. Furthermore, we substantiate this claim by illustrating an example that confirms the 2-approximation ratio and shows the tightness of the result.

\begin{lemma}\label{lm:FF=NF}
If the fixed order is determined by non-increasing slacks, then the output of the next-fit algorithm is identical to that of the first-fit algorithm. 
\end{lemma}
\begin{pf}
Consider an iteration of the first-fit algorithm where $i$ machines are already opened and let $j \in J$ be the job at the beginning of the sorted list of jobs. Then, if job $j$ cannot fit in machine $k$, for all $i = 1,\ldots,k$ we have
\begin{equation}\label{eq:NF=FF}
    L_i + p_j > d_j \Rightarrow L_i > d_j - p_j.
\end{equation}
 Given the assumption that the order is established by non-increasing slacks, for all $q \in J$ with $j < q$, the inequality $d_j - p_j \geq d_q - p_q$ holds. Utilizing this inequality together with inequality \eqref{eq:NF=FF}, we have $L_i +p_q > d_q$, for all $i = 1,\ldots,k$. This implies that while the FF algorithm considers machines $1$ to $k$, if job $j$ cannot fit into any of those machines, then none of the jobs $q$ with $j < q$ can be accommodated in those machines either. This scenario is equivalent to exclusively considering the last machine. Therefore, the lemma holds.\qed
\end{pf}
In the following theorem, we establish that when the order is dictated by non-increasing slacks, both the FF algorithm and, equivalently, the NF algorithm (as shown in Lemma \ref{lm:FF=NF}) attain an approximation ratio of at most~2.

\begin{theorem}\label{th:NF=2}
    First-fit is a 2-approximation algorithm for the case of non-increasing slacks, and this bound is tight.
\end{theorem}
\begin{pf}
For any instance $I$, we prove by induction on $k$ that for all $k \in \mathbb{N}$, if $OPT(I) = k$ then, $FF(I)<2k$.

\textbf{Base Case:  }If OPT fits all of the jobs in one machine then FF can do the same. So, if $OPT(I) =1$, then $FF(I)=1$.

\textbf{Induction Hypothesis:  }Assume that for all $k^{'} \leq k$, if $OPT(I) = k^{'}$ then, $FF(I)<2k^{'}$.

\textbf{Induction Step:  } Assume that $OPT(I) = k + 1$. We use proof by contradiction. Let us assume, to the contrary, that $FF(I) \geq 2k+2$. We assume that machines $i\in \mathcal{M^*}$ are ordered in such a way that 
\begin{align*}
    l^*_1<l^*_2< \ldots<l^*_{k+1}. 
\end{align*}
Thus, the last job in the order (the job with the smallest slack) is scheduled on machine $k+1$ in OPT. Moreover, w.l.o.g. we further assume that $I$ is the smallest such instance in terms of the number of jobs. This means that $l^*_{k+1}$ is the only job on machine $2k+2$ in FF, as otherwise there is a smaller instance by removing all lower priority jobs. Note that by removing such jobs, the number of machines in the optimal schedule does not change because it contradicts the induction hypothesis and FF uses $2k+2$ machines. Thus, $l^*_{k+1} = n$.
\begin{claims}\label{lm:ub1}
For all $i=1,\ldots, 2k+1$,
\begin{equation}
    L^*_{k+1} < L_i + p_{l^*_{k+1}}.
\end{equation}
\end{claims}
\begin{pf}
The assignment of job $l^*_{k+1}$ on machine $k+1$ in OPT implies that $L^*_{k+1} \leq d_{l^*_{k+1}}$. Given that $l^*_{k+1}$ is scheduled on $2k+2$ in FF, it means that it is rejected on all machines $1,\ldots,2k+1$, which implies that $L_i + p_{l^*_{k+1}} > d_{l^*_{k+1}}$ for all $i=1,\ldots, 2k+1$. Using the above two inequalities, we conclude that 
\begin{align*}
    L^*_{k+1} < L_i + p_{l^*_{k+1}}. \tag{$\forall i \in \{1, \ldots, 2k+1\}$}
\end{align*}
\qed
\end{pf}
\begin{claims}\label{lm:ub2}
For all $i<k+1$, we have
\begin{equation}
    L^*_i < L_q + p_{l^*_i},
\end{equation}
for all $q \in \{ 1,\ldots,2i\}$.
\end{claims}
\begin{pf}
Consider $\tau$ as the schedule generated by the FF algorithm, where $\tau(j)$ denotes the machine on which job $j \in J$ is scheduled. Assume that $i<k+1$. We will show that $\tau(l^*_i)\geq 2i+1$ in FF. The result then follows because $L^*_i\leq d_{l^*_i}<L_q+p_{l^*_i}$ as job $l^*_i$ is rejected on machines $q=1,\ldots, 2i$. Suppose, by contradiction, that job $l^*_i$ is scheduled on machine $q\leq 2i$ in FF. Let job $j$ be the first job on machine $\tau(l^*_i)+1$ in FF. Observe from Lemma \ref{lm:FF=NF} that $j>l^*_i.$ Consider the instance $I'$ defined by jobs $j,\ldots,n$. Given that in OPT for $I$, all these jobs are assigned to machines $i+1,\ldots k+1$, we know that $OPT(I')\leq k+1-i$. Moreover, applying FF to $I'$ yields the same number of machines as FF needed for these jobs in $I$, because of Lemma \ref{lm:FF=NF}. Hence, $FF(I')=2k+2-q\geq 2(k+1-i)$. This however contradicts the induction hypothesis of Theorem \ref{th:NF=2}.\qed
\end{pf}
We conclude the proof by using the above two claims
to derive the following contradictory statement.
$$\sum_{i=1}^{k+1}L_i^*< \sum_{q=1}^{2k+2}L_q,$$
where both sides of the inequality should indeed be equal to the sum of processing times of all jobs. Define $M_l=\{q\in\{1,\ldots 2k+2\}\mid l^*_i\in J_{q}\text{ for some }i\in \{1,\ldots,k+1\}\}$. For all $q\in M_l$, we have that $L_q\geq \sum_{l^*_i\in J_q}p_{l^*_i}$ and hence
\begin{equation}\label{inq:matching1}
    \sum_{i=1}^{k+1}p_{l^*_i}\leq \sum_{q\in M_l}L_q.
\end{equation}
As we proved earlier in Claim \ref{lm:ub2}, for each $l^*_i$ where $i\in \{1,\ldots,k\}$, we have $\tau(l^*_i)\geq 2i+1$. To each $l^*_i$ with $i \in \{1,\ldots,k\}$, we assign a machine $q(l^*_i)\in M\setminus M_l$ with $q(l^*_i)\leq 2i$ and $q(l^*_i)\neq q(l^*_v)$ for all $v \in \{1,\ldots,i-1\}$. Such an assignment is feasible, because for each $l^*_i$ with $i \in \{1, \ldots, k\}$, there are $2i$ potential machines. Out of these $2i$ machines, at most $i-1$ machines are in $M_l$ (all jobs $v$ with $v\geq l^*_i$ have $\tau(v)\geq 2i+1$) and at most $i-1$ machines are used for the assignment of a job $l^*_v$ with $v = 1, \ldots, i-1$. This means that at least two candidates are available for $q(l^*_i)$. Moreover, we assign job $l^*_{k+1}$ to a machine $q(l^*_{k+1})\in M\setminus M_l$ with $q(l^*_{k+1})\leq 2k+1$ and $q(l^*_{k+1})\neq q(l^*_v)$ for all $v \in \{1,\ldots,k\}$. Such an assignment is feasible because there are $2k+1$ potential machines. Out of these $2k+1$ machines, at most $k$ machines are in $M_l$, and at most $k$ machines are used for the assignment of a job $l^*_v$ with $v = 1, \ldots, k$. This means that at least one candidate is available for $q(l^*_{k+1})$. Now, for each $i=1,\ldots k+1$, we have by either Claim \ref{lm:ub1} or Claim \ref{lm:ub2} that $L_i^*<L_{q(l^*_i)}+p_{l^*_i}$. Summing over all $i=1,\ldots k+1$ yields
\begin{align*}
 \sum_{i=1}^{k+1}L_i^* < \sum_{i=1}^{k+1}\left[ 
   L_{q(l^*_i)}+ p_{l^*_i} \right] \leq \sum_{q\in M\setminus M_l}L_q+ \sum_{q\in M_l}L_q= \sum_{q=1}^{2k+2}L_q,
\end{align*}
which is a contradiction. By the principle of induction, the claim holds for all $k \in \mathbb{N}$.\qed
\end{pf}

In what follows, we present an example that illustrates the tightness of the bound of~2.
\begin{figure}[t!]
\centering
\includegraphics[scale=0.55]{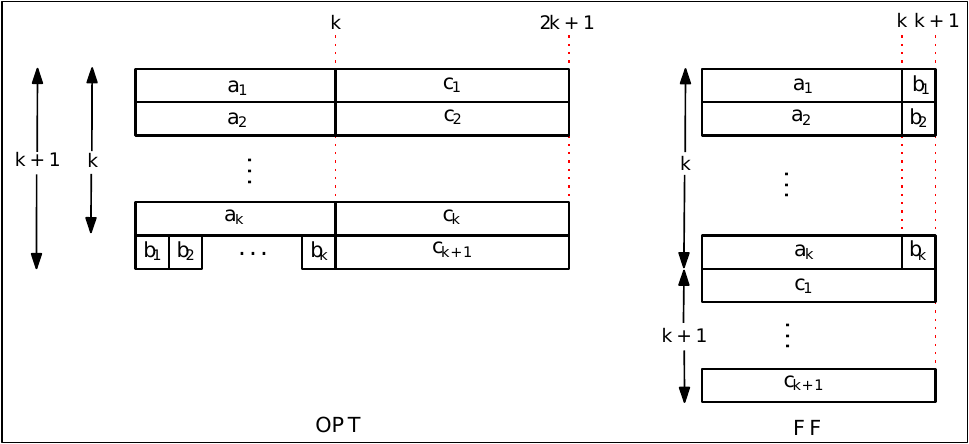}
\caption{An example with the approximation ratio of $(2k+1)/(k+1).$} \label{tightNF}
\end{figure}
\begin{example}\label{ex1}
Consider the following example where $n = 3k+1$ for some $k \ge 1$, featuring three distinct types of jobs
. Specifically, $\mathcal{A} = \{a_1,\ldots,a_k\}$, $\mathcal{B} = \{b_1,\ldots,b_k\}$, and $\mathcal{C} = \{c_1,\ldots,c_{k+1}\}$. In this context, the pair $(k, 2k)=(p_{a_i}, d_{a_i})$ for $a_i \in \mathcal{A}$, the pair $(1, k+1)=(p_{b_i}, d_{b_i})$ for $b_i \in \mathcal{B}$, and the pair $(k+1, 2k+1)=(p_{c_i}, d_{c_i})$ for $c_i \in \mathcal{C}$. Assume the fixed order is given by $a_1 < b_1 < a_2 < b_2 < \cdots < a_k < b_k < c_1 < c_2 < \cdots < c_{k+1}$. Then, the approximation ratio is obtained as $\frac{2k+1}{k+1} < 2$.
\end{example}
\subsection{Other cases}

We present the following theorems for additional cases. Detailed proofs are provided in Appendix~\ref{appendix:A}.

\begin{theorem}
    First-fit is a 2-approximation for the case of non-decreasing slacks, and this bound is tight.
\end{theorem}

\begin{theorem}
    First-fit is a 2-approximation for the case of non-increasing deadlines.
\end{theorem}

\begin{theorem}
   First-fit is a 3/2-approximation when the optimal schedule opens at most two machines.
\end{theorem}

\begin{theorem}
    First-fit is a 2-approximation when the optimal schedule uses at most three machines.
\end{theorem}

\subsection{Logarithmic Approximation Factor for an Arbitrary Order of Jobs}

In this subsection, we establish that the general case of the problem, with an arbitrary fixed order, can be approximated within a factor of \( O(\log n) \). This result is based on the \textit{set cover} problem as described in \cite{vazirani2001approximation}.

\begin{theorem}\label{th:lognApprox}
   The problem $P|r_j = r,p_j, d_j, \pi^{*}|m$ can be approximated within a factor of $O(\log n)$.
\end{theorem}
\begin{pf}
The set cover problem is formulated as follows: Let \(\mathcal{U}\) be a universe of \(n\) elements, and let \(\mathcal{S} = \{S_1, S_2, \ldots, S_k\}\) be a collection of subsets of \(\mathcal{U}\). Each subset \(S_i \in \mathcal{S}\) is associated with a cost \(c(S_i) \in \mathbb{Q}^+\). The objective is to find a subcollection of \(\mathcal{S}\) that covers all elements of \(\mathcal{U}\) while minimizing the total cost. A greedy method provides an \(O(\log n)\)-approximation for this problem, as shown in \cite{vazirani2001approximation}. This algorithm iteratively selects the most \textit{cost-effective} subset which is the subset that minimizes the cost per uncovered element, removes the elements covered by the selected subset, and repeats the process until all elements in \(\mathcal{U}\) are covered. 

We define \(\mathcal{S}\) as the collection of subsets of jobs \(S_i\), where each subset contains the jobs that can be scheduled on a single machine in the given fixed order without violating any deadlines. Let \(\bar{J}\) represent the set of unscheduled jobs, initially equal to \(J\). While \(\bar{J} \neq \emptyset\), perform the following steps
\begin{itemize}
    \item Identify a subset \(S_i\), where \(i \in \{1, \ldots, |\mathcal{S}|\}\), that contains the largest number of unscheduled jobs. The subset \(S_i\) must satisfy the condition that all jobs in \(S_i\) can be executed on a single machine in the given fixed order without violating any deadlines.
    \item Remove all jobs \(j \in S_i\) from \(\bar{J}\).
\end{itemize}
According to \cite{vazirani2001approximation}, the described greedy method results in a $O(\log n)$ approximation factor for the problem $P|r_j = r,p_j, d_j, \pi^{*}|m$. Given that there are an exponential number of subsets, we present a dynamic programming approach that runs in polynomial time to identify a subset \( S_i \subseteq \mathcal{S} \) that contains the maximum number of unscheduled jobs that can be successfully scheduled on a single machine without violating their respective deadlines. Let \( D(i, k) \) denote the smallest possible completion time for scheduling \( k \) jobs selected from the set \( \{1, \ldots, i\} \) on a single machine, such that no job violates its deadline. The recurrence relation for \( D(i, k) \) is defined as follows

\textbf{\textit{Case (I):}} Job \( i \) is not included in the solution.
If job \( i \) is not a part of the solution, the shortest completion time for \( k \) jobs remains the same as for \( k \) jobs chosen from the subset \( \{1, \ldots, i-1\} \). So, $D(i, k) = D(i - 1, k)$.

\textbf{\textit{Case (II):}} Job \( i \) is included in the solution.
If job \( i \) is included in the solution, then \( k - 1 \) jobs must be selected from the subset \( \{1, \ldots, i-1\} \). The completion time for these \( k - 1 \) jobs must allow job \( i \) to be scheduled within its deadline \( d_i \). In this case, if $D(i - 1, k - 1) + p_i \leq d_i$ we have
\[
D(i, k) = D(i - 1, k - 1) + p_i.
\]
Therefore,
\[
D(i, k) = 
\begin{cases}
  \min \big( D(i - 1, k), D(i - 1, k - 1) + p_i \big),  & \text{if } D(i - 1, k - 1) + p_i \leq d_i\\
  D(i - 1, k), & \text{otherwise}
\end{cases}
\]
The boundary conditions are \( D(0, 0) = 0 \), \( D(i, 0) = 0\text{, for all } i \), and \( D(0, k) = +\infty \, \text{, for all } k > 0 \). Using this dynamic programming approach, we search for the maximum $k$ for which $D(n, k) < +\infty,$ which can be done in polynomial time $O(n^2)$. Given that we need to run the dynamic program at most $O(n)$ times, the total running time is $O(n^3)$.
\qed
\end{pf}

\section{Conclusion}
This paper studied the SRDM problem with an imposed fixed order and common release dates. We showed that the greedy algorithm next-fit has an unbounded approximation ratio for this problem. We proved that the first-fit algorithm solves the restricted case of unit processing times in polynomial time. When the order aligns with non-increasing slacks, non-decreasing slacks, non-increasing deadlines, or opens at most 3 machines, we proved that the greedy first-fit algorithm guarantees a 2-approximation. Our results seem to suggest that the first-fit algorithm approximates the optimal solution within a factor of two for arbitrary orders, but we have not been able to give a formal proof. Further research could also examine other algorithms as potential approximation methods to determine if they offer better bounds.

\newpage
\appendix
\section {Other Cases}\label{appendix:A}
\subsection{Non-Decreasing Slacks}
In this subsection, we prove that the approximation ratio of the first-fit algorithm is 2 for the case where the fixed order matches the non-decreasing slacks.

\setcounter{theorem}{2}

\begin{theorem}\label{th:FF=2-NDS}
    First-fit is a 2-approximation for the case of non-decreasing slacks. Furthermore, this bound is tight.
\end{theorem}
\begin{pf}
For any instance $I$, we prove by induction on $k$ that for all $k \in \mathbb{N}$, if $OPT(I) = k$ then, $FF(I)<2k$.

\textbf{Base Case:  }If OPT fits all of the jobs in one machine then FF can do the same. Therefore, if $OPT(I) =1$, then $FF(I)=1$.

\textbf{Induction Hypothesis:  }Assume that for all $k^{'} \leq k$, if $OPT(I) = k^{'}$ then, $FF(I)<2k^{'}$.

\textbf{Induction Step:  } Assume that $OPT(I) = k + 1$. We use a proof by contradiction. To the contrary, let us assume that $FF(I) \geq 2k+2$. Likewise, in Theorem \ref{th:NF=2}, we assume that $I$ is the smallest such instance by means of the number of jobs and that machines $i\in \mathcal{M^*}$ are ordered in such a way that $ l^*_1<l^*_2< \ldots<l^*_{k+1}$. Thus, the last job in the order (the job with the highest slack value) is scheduled on machine $k+1$ in OPT.

\begin{claims}\label{lm:ub2-NDS}
For all $i \leq k+1$, we have
\begin{equation}
    L^*_i < L_q + p_{l^*_i},
\end{equation}
for all $q\in \{1, \ldots, 2k+1\}$.
\end{claims}
\begin{pf}
Since $l^*_{k+1}$ is rejected on machines $1,\ldots, 2k+1$, we know that for all $i=1,\ldots, 2k+1$, we have $d_{l^*_{k+1}} - p_{l^*_{k+1}} < L_i$. Moreover, based on the assumption we know that $l^*_{k+1}$ has the largest slack, i.e., $d_j - p_j \leq d_{l^*_{k+1}} - p_{l^*_{k+1}}$ for all $j \in J$. Furthermore, we have that $L^*_i - p_{l^*_i} \leq d_{l^*_i} - p_{l^*_i}$. Utilizing the previous inequalities for all $i=1,\ldots,k+1$, provides us with

\begin{align*}
    L^*_i < L_q + p_{l^*_i}. \tag{$\forall q\in \{1, \ldots, 2k+1\}$}
\end{align*}
\qed 
\end{pf}
We can now use the above claim to derive the contradictory statement that $\sum_{i=1}^{k+1}L_i^*< \sum_{q=1}^{2k+2}L_q$, where both sides of the inequality should indeed be equal to the sum of processing times of all jobs.
 
Define $M_l=\{q\in\{1,\ldots 2k+2\}\mid l^*_i\in J_{q}\text{ for some }i\in \{1,\ldots,k+1\}\}$. For all $q\in M_l$, we have that $L_q\geq \sum_{l^*_i\in J_q}p_{l^*_i}$ and hence, we have that 
\begin{equation}\label{inq:matching11}
    \sum_{i=1}^{k+1}p_{l^*_i}\leq \sum_{q\in M_l}L_q.
\end{equation}

To each $l^*_i$ with $i \in \{1,\ldots,k+1\}$, we assign a machine $q(l^*_i)\in M\setminus M_l$ with $q(l^*_i)\neq q(l^*_v)$ for all $v \in \{1,\ldots,k+1\}$ and $l^*_i \neq l^*_v$. Such an assignment is feasible, because based on Claim \ref{lm:ub2-NDS}, for each $l^*_i$ with $i \in \{1, \ldots, k+1\}$, there are $2k+1$ potential machines. Out of these $2k+1$ machines, at most $k$ machines are in $M_l$ (note that job $l^*_{k+1}$ is alone in machine $2k+2$, thus $2k+2 \in M_l$), and at most $k+1$ machines are used for the assignment of a job $l^*_v$ with $v = 1, \ldots, k+1$. This means that enough candidates are available for $q(l^*_i)$.

Now, for each $i=1,\ldots k+1$, we have by Claim \ref{lm:ub2-NDS} that $L_i^*<L_{q(l^*_i)}+p_{l^*_i}$. Summing over all $i=1,\ldots k+1$ yields
\begin{align*}
 \sum_{i=1}^{k+1}L_i^* < \sum_{i=1}^{k+1}\left[ 
   L_{q(l^*_i)}+ p_{l^*_i} \right] \leq \sum_{q\in M\setminus M_l}L_q+ \sum_{q\in M_l}L_q= \sum_{q=1}^{2k+2}L_q,
\end{align*}
which is a contradiction.

By the principle of induction, the claim holds for all $k \in \mathbb{N}$.
\qed
\end{pf}

\begin{rmk}
In Example \ref{ex1}, the slack of all the jobs is equal. Therefore, it can be used as a lower bound example to show the tightness of ratio 2 in the case of non-decreasing slacks.
\end{rmk}

\subsection{Non-Increasing Deadlines}
In this subsection, we prove that the approximation ratio of the first-fit algorithm is 2 for the case where the fixed order matches the non-increasing order of the deadlines.

\begin{theorem}\label{th:FF=2-NID}
    First-fit is a 2-approximation algorithm for the case of non-increasing deadlines.
\end{theorem}
\begin{pf}
For any instance $I$, we prove by induction on $k$ that for all $k \in \mathbb{N}$, if $OPT(I) = k$ then, $FF(I)<2k$.

\textbf{Base Case:  }If OPT fits all of the jobs in one machine then FF can do the same. Therefore, if $OPT(I) =1$, then $FF(I)=1$.

\textbf{Induction Hypothesis:  }Assume that for all $k^{'} \leq k$, if $OPT(I) = k^{'}$ then, $FF(I)<2k^{'}$.

\textbf{Induction Step:  } Assume that $OPT(I) = k + 1$. We use a proof by contradiction. Let us assume, to the contrary, that $FF(I) \geq 2k+2$. Moreover, as in Theorem \ref{th:NF=2}, we assume that $I$ is the smallest such instance by means of the number of jobs.

For all $i=1,\ldots,k+1$, define $E_i=\{j\in J_i^*\mid \tau(j) =  \max (\tau(k)) \text{ for all } k\in J_i^*\}$ and let $e^*_i \in E_i$ be the job with $e^*_i\geq j$ for all $j \in E_i$. We further assume that machines $i\in \mathcal{M^*}$ are ordered in such a way that 

\begin{align*}
    \tau(e^*_1) \leq \tau(e^*_2) \leq \ldots \leq \tau(e^*_{k+1}). 
\end{align*}

Hence, the last job in the order (the job with the smallest deadline) is scheduled on machine $k+1 \in \mathcal{M^*}$. Notice that $e^*_{k+1} = n$.
\begin{figure}[t!]
\centering
\includegraphics[scale=0.55]{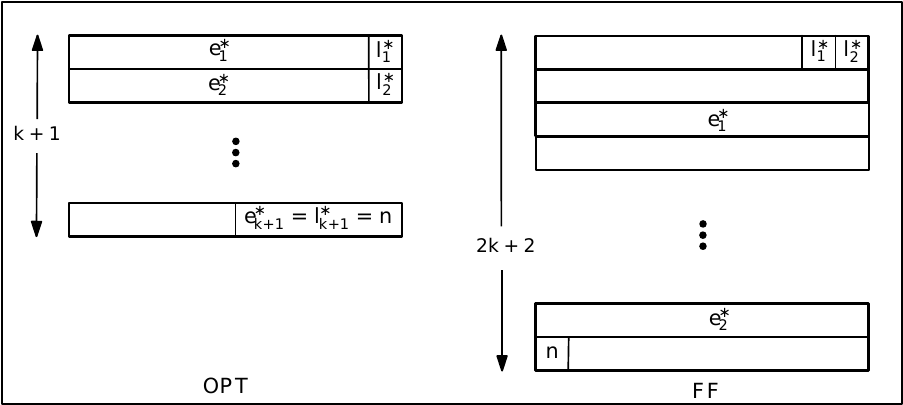}
\caption{An example of the placement of jobs $e^*_i$ in the OPT and FF schedules.} \label{edefinition}
\end{figure}

\begin{claims}
For all $i < k+1$, we have
\begin{equation}
    L^*_i < L_q + p_{e^*_i},
\end{equation}
for all $q \in \{ 1,\ldots,2i\}$.
\end{claims}
\begin{pf}
Assume that $i<k+1$. We will show that $\tau(e^*_i)\geq 2i+1$ in FF. The result then follows because $L^*_i\leq d_{l^*_i}\leq d_{e^*_i} <L_q+p_{e^*_i}$ as job $e^*_i$ is rejected on machines $q=1,\ldots, 2i$. 

Suppose, by contradiction, that job $e^*_i$ is scheduled on a machine $q\leq 2i$ in FF. Based on the definition of job $e^*_i$, we know that $\tau(j) \leq q$ for all jobs $j \in J^*_i$. Moreover, based on our earlier assumption on the order of the machines in the optimal schedule, we know that $\tau(j) \leq q$ for all $j \in \cup^{s=1}_{i-1}J^*_s$. Consider the instance $I'$ defined by jobs $j \in \cup^{s=i+1}_{k+1}J^*_s$. Given that in the optimal schedule for $I$, all these jobs are assigned to machines $i+1,\ldots k+1$, we know that $OPT(I') = k+1-i$. Moreover, applying FF to $I'$ yields at least the same number of machines as FF needed for these jobs in $I$. Hence, $FF(I')=2k+2-q\geq 2(k+1-i)$. This however contradicts the induction hypothesis. 
\qed 
\end{pf}

Also, using the same argument as in Claim \ref{lm:ub1}, it can be proved that for all $i=1,\ldots, 2k+1$, we have $ L^*_{k+1} < L_i + p_{e^*_{k+1}}$ (note that $e^*_{k+1} = l^*_{k+1} = n$). By utilizing this inequality and the above claim
we can conclude the proof by driving the contradictory statement of $\sum_{i=1}^{k+1}L_i^*< \sum_{q=1}^{2k+2}L_q$, where both sides of the inequality should indeed be equal to the sum of processing times of all jobs.
 
Define $M_e=\{q\in\{1,\ldots, 2k+2\}\mid e^*_i\in J_{q}\text{ for some }i\in \{1,\ldots,k+1\}\}$. For all $q\in M_e$, we have that $L_q\geq \sum_{e^*_i\in J_q}p_{e^*_i}$ and hence, we have that 
\begin{equation}
    \sum_{i=1}^{k+1}p_{e^*_i}\leq \sum_{q\in M_e}L_q.
\end{equation}

By the claim we have that 
for each $e^*_i$ where $i\in \{1,\ldots,k\}$, we have $\tau(e^*_i)\geq 2i+1$. To each $e^*_i$ with $i \in \{1,\ldots,k\}$, we assign a machine $q(e^*_i)\in M\setminus M_e$ with $q(e^*_i)\leq 2i$ and $q(e^*_i)\neq q(e^*_v)$ for all $v \in \{1,\ldots,i-1\}$. Such an assignment is feasible, because for each $e^*_i$ with $i \in \{1, \ldots, k\}$, there are $2i$ potential machines. Out of these $2i$ machines, at most $i-1$ machines are in $M_e$ (all jobs $e^*_v$ with $v > i$ have $\tau(e^*)\geq 2i+1$) and at most $i-1$ machines are used for the assignment of a job $e^*_v$ with $v = 1, \ldots, i-1$. This means at least two candidates are available for $q(e^*_i)$. Moreover, we assign job $e^*_{k+1}$ to a machine $q(e^*_{k+1})\in M\setminus M_e$ with $q(e^*_{k+1})\leq 2k+1$ and $q(e^*_{k+1})\neq q(e^*_v)$ for all $v \in \{1,\ldots,k\}$. Such an assignment is feasible because there are $2k+1$ potential machines. Out of these $2k+1$ machines, at most $k$ machines are in $M_e$, and at most $k$ machines are used for the assignment of a job $e^*_v$ with $v = 1, \ldots, k$. This means at least one candidate is available for $q(e^*_{k+1})$.

As mentioned before, using Claim \ref{lm:ub1}, it can be proved that for all $i=1,\ldots, 2k+1$, we have $ L^*_{k+1} < L_i + p_{e^*_{k+1}}$. Now, for each $i=1,\ldots k+1$, we have by either the previous inequality or by the above claim
that $L_i^*<L_{q(e^*_i)}+p_{e^*_i}$. Summing over all $i=1,\ldots k+1$ yields
\begin{align*}
 \sum_{i=1}^{k+1}L_i^* < \sum_{i=1}^{k+1}\left[ 
   L_{q(e^*_i)}+ p_{e^*_i} \right] \leq \sum_{q\in M\setminus M_e}L_q+ \sum_{q\in M_e}L_q= \sum_{q=1}^{2k+2}L_q,
\end{align*}
which is a contradiction.

By the principle of induction, the claim holds for all $k \in \mathbb{N}$.
\qed
\end{pf}

\subsection{Analyses of the Approximation Factors for a Limited Number of Machines}

In this subsection, we allow for an arbitrary fixed order. We demonstrate constant approximation factors of 3/2 and 2 when the number of machines used in OPT is 2 and 3, respectively. These analyses highlight how the complexity increases when proving a constant approximation factor for the first-fit algorithm.

\begin{theorem}\label{th:FF=2-4}
   First-fit is a 3/2-approximation when the optimal schedule opens at most two machines.
\end{theorem}
\begin{pf}
 Consider any instance $I$. Observe that if $OPT(I)$ = 1, then $FF(I) =1$.

 In the following, we prove that if $OPT(I) = 2$ then, $FF(I) < 4$. Assume by contradiction that for some instance $I^{'}$, $OPT(I^{'})=2$ and $FF(I^{'}) = 4$ (as before we consider $I^{'}$ to be the smallest such instance in terms of the number of jobs).

 Assume w.l.o.g. that job $n$ is scheduled in the second machine in the optimal schedule. We have that $L^*_2 \leq d_n$. We also know that job $n$ is rejected from the first three machines in the FF algorithm. Thus, $d_n < L_i + p_n$ with $i = 1,2,3$. Hence, for $i \in \{1,2,3\}$ we have
 \begin{equation}\label{eq:241}
         L^*_2 < L_i + p_n.
 \end{equation}

 \begin{figure}
\centering
\includegraphics[scale=0.55]{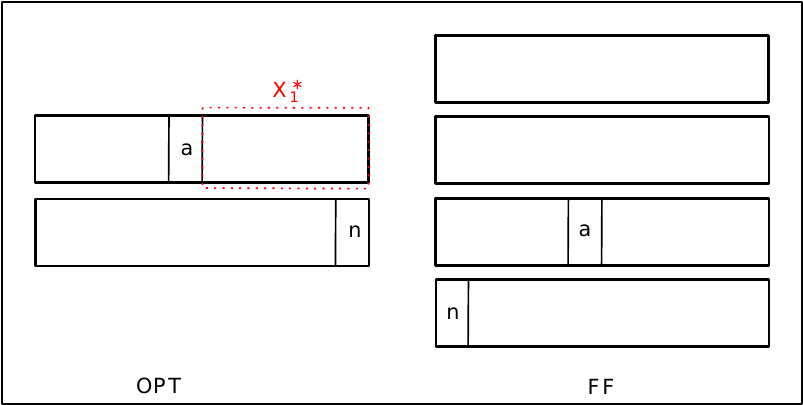}
\caption{Instance $I^{'}$ where $OPT(I^{'})=2$ and $FF(I^{'}) = 4$.} \label{24}
\end{figure}

 Observe that $J_3 \setminus J_2^* \neq \emptyset$, as otherwise, the last job could have fit in the third machine. Therefore, there should be a job $a \in J^*_1\cap J_3$. Let $a \in J^*_1$ be such a job with the lowest possible priority. Denote also by $X^*_1$ the set of all jobs $j \in J^*_1$ with $j > a$ (see Figure \ref{24} for an illustration). Note that set $X^*_1$ can be empty. Below, we demonstrate that all elements $b \in X^*_1$ are allocated to the first machine by the FF algorithm.

\begin{lemma}\label{lm:24}
First-fit schedules all the jobs $b\in X^*_1$ in the first machine.
\end{lemma}
\begin{pf}
 We prove this by contradiction. First, consider inequality (\ref{eq:241}) with $i = 3$. We have
 \begin{equation}\label{eq:242}
 L^*_2 < L_3 + p_n\Leftrightarrow
 \sum_{\substack{j \in J^*_2 \\ j \in J_1}}p_j + \sum_{\substack{j \in J^*_2 \\ j \in J_2}}p_j+ \sum_{\substack{j \in J^*_2 \\ j \in J_3}}p_j  < \sum_{\substack{j \in J_3 \\ j \in J^*_1}}p_j + \sum_{\substack{j \in J_3 \\ j \in J^*_2}}p_j\Leftrightarrow
        \sum_{\substack{j \in J^*_2 \\ j \in J_1}}p_j +  \sum_{\substack{j \in J^*_2 \\ j \in J_2}}p_j<  \sum_{\substack{j \in J_3 \\ j \in J^*_1\\j \neq a}}p_j + p_a. 
\end{equation}
 
 Now, assume by contradiction that a job $b \in X^*_1$ is not scheduled in the first machine by the FF algorithm. Note that job $b$ cannot reside in the third machine due to $b > a$ (as per the definition, job $a$ should hold the lowest priority in the third machine, belonging to $J^*_1$). Thus, job $b$ should be scheduled in the second machine.

The optimal schedule fits job $b$ in the first machine. We get
\begin{equation}\label{eq:243}
    \sum_{\substack{j \in J^*_1 \\ j < b \\j \neq a}}p_j  + p_a + p_b \leq d_b.
\end{equation}
Also, job $b$ is rejected from the first machine by the FF algorithm. We get
\begin{equation}\label{eq:244}
    \sum_{\substack{j \in J^*_1 \\ j \in J_1 \\j < b}}p_j + \sum_{\substack{j \in J^*_2 \\ j \in J_1 \\j < b}}p_j + p_b > d_b.
\end{equation}
By combining inequalities (\ref{eq:243}) and (\ref{eq:244}), we have
\begin{equation}\label{eq:245}
\sum_{\substack{j \in J^*_1 \\ j < b \\j \neq a}}p_j  + p_a  <  \sum_{\substack{j \in J^*_1 \\ j \in J_1 \\j < b}}p_j + \sum_{\substack{j \in J^*_2 \\ j \in J_1 \\j < b}}p_j\Leftrightarrow \sum_{\substack{j \in J^*_1 \\ j \in J_2 \\j < b}}p_j +\sum_{\substack{j \in J^*_1 \\ j \in J_3 \\j < b \\j \neq a}}p_j  + p_a  < \sum_{\substack{j \in J^*_2 \\ j \in J_1 \\j < b}}p_j\Leftrightarrow \sum_{\substack{j \in J^*_1 \\ j \in J_2 \\j < b}}p_j +\sum_{\substack{j \in J^*_1 \\ j \in J_3\\j \neq a}}p_j  + p_a  < \sum_{\substack{j \in J^*_2 \\ j \in J_1 \\j < b}}p_j,
\end{equation}
where the last equivalence follows because all jobs in $X_1^*$ are in $J_1$ or $J_2$.

Since the processing times are positive, it is easy to verify that inequalities (\ref{eq:242}) and (\ref{eq:245}) contradict. Therefore, the proof holds.
\qed 
\end{pf}
The optimal schedule fits job $a$ in the first machine. We get
\begin{equation}\label{eq:246}
    L^*_1 -  \sum_{\substack{j \in X_1^*}}p_j \leq d_a
\end{equation}
Also, job $a$ is rejected from the first machine by the FF algorithm. We get
\begin{equation}\label{eq:247}
    L_1 -  \sum_{\substack{j \in J_1 \\ j > a}}p_j + p_a > d_a
\end{equation}
By combining inequalities (\ref{eq:246}) and (\ref{eq:247}), we have
\begin{equation}\label{eq:248}
    L^*_1 -  \sum_{\substack{j \in X^*_1 }}p_j < L_1 -  \sum_{\substack{j \in J_1 \\ j > a}}p_j + p_a.
\end{equation}
Based on Lemma \ref{lm:24}, we know that all jobs $b \in X^*_1$ belong to $J_1$. This means that $\sum\limits_{\substack{j \in J_1 \\ j > a}} p_j$ contains $\sum\limits_{\substack{j \in X^*_1}} p_j$. Thus
\begin{equation}\label{eq:249}
    L^*_1 < L_1 + p_a.
\end{equation}
Consider inequality (\ref{eq:241}) with $i = 2$. By summing up inequalities (\ref{eq:241}) and (\ref{eq:249}), we obtain
\begin{alignat*}{2}
L^*_1 + L^*_2 & <  L_1 + p_a + L_2 + p_n \\
 & < L_1 + L_2 + L_3 + L4, 
\end{alignat*}
where second inequality is derived from the fact that jobs $a$ and $n$ belong to $J_3$ and $J_4$, respectively. Thus, their processing times cannot exceed the load of their respective machines. This contradicts the fact that the total loads in both the optimal schedule and the FF algorithm schedule must be equal. 
\qed 
\end{pf}

Now, we demonstrate that the FF algorithm achieves a 2-approximation factor when the optimal schedule employs three machines.

\begin{theorem}\label{th:FF=last}
    First-fit is a 2-approximation when the optimal schedule uses at most three machines.
\end{theorem}
\begin{pf}
For any instance $I$, we prove that if $OPT(I) = 3$ then, $FF(I) < 7$. Observe that if $OPT(I)$ = 1, then $FF(I) =1$, and when $OPT(I)$ = 2, based on Theorem \ref{th:FF=2-4}, $FF(I) < 4$. Assume by contradiction that there is an instance $I$ with $OPT(I)=3$ and $FF(I) = 7$. As before we assume that $I$ is an instance with those properties having the fewest number of jobs. In particular, the lowest priority job $n$ is the only job assigned to the last opened machine when the FF algorithm terminates. 

For each machine $i \in \{ 1,2,3\}$ in the optimal solution, we define a critical job $c_i^*$ to be the lowest priority job which is scheduled on machine $i$ in OPT and on machines 4 to 7 by the FF algorithm, $c_i^* =\max \{ j:  \tau^*(j) = i, \tau(j) \ge 4 \}.$

We assume that the machines in OPT are ordered such that $c_1^* < c_2^* < c_3^*$. Clearly, $c_3^* = n$, and for $i = 1,2$ it needs to be shown that $c_i^*$ exists. 
Assume to the contrary that $\tau(j) \le 3$ for all $j$ with $\tau^*(j)=1$. This means that for any job $j$ with $\tau(j) \in \{4,\ldots, 7\}$, we have $\tau^*(j) \in \{2,3\}$. If we now consider the instance with jobs only in the set $J'=\{ j: 4\le \tau(j) \le 7\}$,
then FF will schedule the jobs in $J'$ just like it would for the instance $I$ on four machines, while those jobs are scheduled on two machines in the optimal solution for $I$. This contradicts Theorem~\ref{th:FF=2-4}. The same argument shows that $c_2^*$ exists.

We will now derive the following inequality
 \begin{equation}\label{eq:370}
\sum_{i=1}^{3}L_i^*< \sum_{i=1}^{7}L_i,
\end{equation}

where $L_i^*$ is the completion time of machine $i$ in the optimal solution and $L_i$ is the completion time of machine $i$ in the FF solution. Both sides of the inequality are equal to the sum of processing times of all jobs and hence this contradicts our initial assumption.

First of all, since job $c_3^* = n$ is scheduled on machine 3 in the optimal solution and rejected from the first six machines by the FF algorithm, we have that
 \begin{equation}\label{eq:371}
         L^*_3 \le d_n < L_i + p_n
 \end{equation}
for every  $i \in \{1,\ldots,6\}$. We now consider two cases.

\textbf{\textit{Case (I):}} there exists a job $j$ such that $c_1^* < j < c_2^*$, $\tau^*(j) = 1$ and $ \tau(j) \in \{2,3\}$. If this is the case, we replace $c_1^*$ with the highest indexed, i.e. lowest priority, job $j$ that satisfies the above three properties.

Let us assume that $\tau(c_1^*) =2 $ and that $\tau(c_2^*)=4$. The cases when $\tau(c_1^*) = 3 $ or when $\tau(c_2^*) \in \{5,6\}$ can be handled similarly by renaming.
Since $c_1^*$ is scheduled on machine 1 in the optimal solution and rejected from machine 1 by the first algorithm, we obtain

 \begin{equation}\label{eq:372}
L_1^* - \sum_{\substack{j > c_1^* \\ \tau^*(j) = 1}} p_j \le 
d_{c_1^*} < L_1 -  \sum_{\substack{j > c_1^* \\ \tau(j) = 1}} p_j + p_{c_1^*}. 
 \end{equation}

Similarly, since  $c_2^*$ is scheduled on machine 2 in the optimal solution and rejected from machine 3 by the first algorithm, we obtain

 \begin{equation}\label{eq:373}
L_2^* - \sum_{\substack{j > c_2^* \\ \tau^*(j) = 2}} p_j \le 
d_{c_2^*} < L_3 -  \sum_{\substack{j > c_2^* \\ \tau(j) = 3}} p_j + p_{c_2^*}. 
 \end{equation}

When we combine equations \eqref{eq:371} (with $i=5$), \eqref{eq:372} and \eqref{eq:373}, we find that

$$L_1^* + L_2^* + L_3^* < L_1 + L_3 + L_5 + p_{c_1^*} + p_{c_2^*} + p_{c_3^*} + \sum_{\substack{j > c_1^* \\ \tau^*(j) = 1}} p_j  + \sum_{\substack{j > c_2^* \\ \tau^*(j) = 2}} p_j - \sum_{\substack{j > c_1^* \\ \tau(j) = 1}} p_j  - \sum_{\substack{j > c_2^* \\ \tau(j) = 3}} p_j.$$

Since $\tau(c_2^*) = 4$ and $\tau(c_3^*) = 7$, we also  have that $p_{c_2^*} \le L_4$ and $p_{c_3^*} \le L_7$. 
Now consider the positive $p_j$ terms in the first sum on the right-hand side of the above inequality. 
By the choice of $c_1^*$, if $j > c_1^*$ and $\tau^*(j)=1$, then either $\tau(j)=1$, in which case there is a corresponding negative term in the third sum, or $j > c_2^*$. In that case, there is a corresponding negative $p_j$ term in the last sum, or $\tau(j)=2$. Similarly, for the positive $p_j$ terms in the second sum it holds that $\tau(j) \le 3$ and there are either corresponding negative terms in the third or last sum, or $\tau(j)=2$.
Hence, $p_{c_1^*}$ plus the last four sums can be bounded from above by $L_2$. Hence, we obtain the desired contradiction \eqref{eq:370} for the first case

$$L_1^* + L_2^* + L_3^* < L_1 + L_3 + L_5 + L_4 + L_7 + L_2 \le \sum_{i=1}^{7}L_i.$$ 

\textbf{\textit{Case (II):}} In this case there is no $j$ such that $c_1^* < j < c_2^*$, $\tau^*(j) = 1$ and $ \tau(j) \in \{2,3\}$, i.e. $\tau(j)=1$ for all such $j$. As before, we have that 

 \begin{equation}\label{eq:374}
L_1^* - \sum_{\substack{j > c_1^* \\ \tau^*(j) = 1}} p_j \le 
d_{c_1^*} < L_1 -  \sum_{\substack{j > c_1^* \\ \tau(j) = 1}} p_j + p_{c_1^*} 
 \end{equation}

 and 
 
 \begin{equation}\label{eq:375}
L_2^* - \sum_{\substack{j > c_2^* \\ \tau^*(j) = 2}} p_j \le 
d_{c_2^*} < L_2 -  \sum_{\substack{j > c_2^* \\ \tau(j) = 2}} p_j + p_{c_2^*},
 \end{equation}

 since $c_2^*$ is rejected from machine 2 by the FF algorithm.

We now use \eqref{eq:371} with that value of $4\le i \le 6$ such that $\tau(c_1^*) \neq i$ and $\tau(c_2^*) \neq i$, e.g. if $\tau(c_1^*) = 4$ and $\tau(c_2^*) =5$ we use $i=6$. Together with \eqref{eq:374} and \eqref{eq:375}, we get that

$$L_1^* + L_2^* + L_3^* < L_1 + L_2 + L_6 + p_{c_1^*} + p_{c_2^*} + p_{c_3^*} + \sum_{\substack{j > c_1^* \\ \tau^*(j) = 1}} p_j  + \sum_{\substack{j > c_2^* \\ \tau^*(j) = 2}} p_j - \sum_{\substack{j > c_1^* \\ \tau(j) = 1}} p_j  - \sum_{\substack{j > c_2^* \\ \tau(j) = 2}} p_j.$$
 
Again, the positive terms $p_j$ in the first two sums either have a corresponding negative term or $\tau(j)=3$ due to the choice of $c_1^*$ and $c_2^*$. Since $p_{c_1^*} + p_{c_2^*} + p_{c_3^*} \le L_4 + L_5 + L_7$, we finally get 

$$L_1^* + L_2^* + L_3^* < L_1 + L_2 + L_6 + L_4 + L_5 +  L_7 + L_3 \le \sum_{i=1}^{7}L_i.$$

This contradicts \eqref{eq:370} also in this case and completes the proof.

\qed
\end{pf}


\end{document}